\documentclass[doublecol]{epl2} 

\usepackage{color}
\usepackage{graphicx}
\usepackage{amsmath}

\title{The shape and mechanics of curved fold origami structures}

\author{Marcelo A. Dias\inst{1,2}\footnote{\email{marcelo\underline{ }dias@brown.edu}} \and Christian D. Santangelo\inst{1}\footnote{\email{csantang@physics.umass.edu}}}
\shortauthor{M. A. Dias \etal}

\institute{                    
  \inst{1} Department of Physics, University of Massachusetts Amherst, Amherst, Massachusetts 01003\\
  \inst{2} School of Engineering, Brown University, Providence, Rhode Island 02912
}

\abstract{
We develop recursion equations to describe the three-dimensional shape of a sheet upon which a series of concentric curved folds have been inscribed. In the case of no stretching outside the fold, the three-dimensional shape of a single fold prescribes the shape of the entire origami structure. To better explore these structures, we derive continuum equations, valid in the limit of vanishing spacing between folds, to describe the smooth surface intersecting all the mountain folds. We find that this surface has negative Gaussian curvature with magnitude equal to the square of the fold's torsion. A series of open folds with constant fold angle generate a helicoid.}

\begin{document}

\maketitle

\section{Introduction}

In the hands of an artist, a single sheet of paper can be folded into a seemingly endless variety of three-dimensional, and strikingly beautiful, forms. Among the various fold patterns, structures built from curved folds, first folded in the late 1920's by a student at the Bauhaus \cite{Wingler69}, stand out as being particularly stunning \cite{Huffman76,DuncanDuncan82,FuchsTabachnikov99,PottmannWallner2001}. The mechanics of these structures arises, at least in part, from the interaction between the avoidance of in-plane stretching within the sheet and the three-dimensional geometry \cite{Dias2012}. Besides its own artistic value, origami has, for many years, shown itself to be a valuable tool to approach problems in architecture, structural mechanics, and engineering \cite{Engel1997,Miura1980,demaine1,Schenk2011,Seffen2012}. This new framework of curved pleated structures has already been demonstrated to be very useful for designing materials with new mechanical properties \cite{Seffen2012} ({\emph e.g.} anisotropic response to external forces and structural stiffness). These techniques could also potentially serve as a platform for design at many different scales.

\begin{figure}
\includegraphics[width=3.25in]{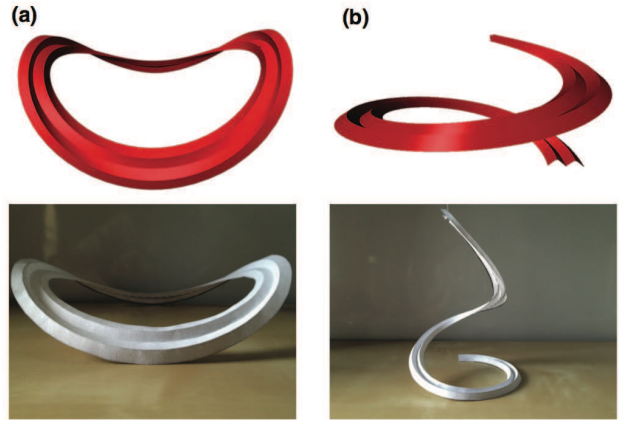}
\caption{\label{fig:multi} A curved-fold origami structure with three prescribed concentric circles that is either (a) closed or (b) open (non-constant curvatures). The top figures demonstrate computational origami, based on three concentric, circular folds of fixed spacing, $w \kappa^1_g=0.1$. {\bf(a)}: the inner fold has curvature $2 \kappa_g^{1}/\sqrt{3}$ and torsion $\tau^1(s) = 1.26 \sin(2 s)$, chosen so that the origami is closed. {\bf(b)}: the initial curvature is $\kappa^1(s)=1.2\kappa_g^1$ and torsion $\tau^1(s)=0.2\kappa_g^1$.} 
\end{figure}

In this article, we will explore the shapes adopted by multiple, circular curved folds such as those shown in Fig. \ref{fig:multi}. In our analysis, we shall assume that the sheets from which the folds are made resist stretching between the creases, which implies that their shapes can be approximated to developable surfaces (surfaces with zero Gaussian curvature \cite{Spivak1999}). Since it is not clear that this is always possible (a fold made from nested squares is not foldable in this strict mathematical sense \cite{Demaine11}), we explore the possible presence of singularities within the sheets. In fact, our findings suggest that a sufficiently narrow spacing between folds is generically sufficient to prevent singularities from appearing. We show that nested helices are a solution for open concentric circular folds, such as in Fig. \ref{fig:multi}{\bf b}. Finally we consider the asymptotic shape of the surface that passes through all the mountain (or valley) folds in the limit that the number of folds diverges. The nested helix solution, in this limit, becomes a portion of a helicoid.

\section{Geometry of multiple folds}

\begin{figure}
\includegraphics[width=3.25in]{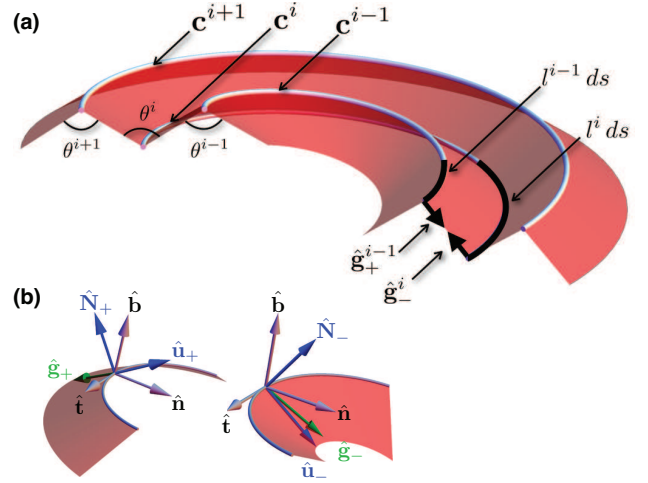}
\caption{\label{fig:multi2}  {\bf(a)}: A schematic representation of the geometry of the generators in the multiple folded configuration. {\bf(b)}:  A mountain fold split along the crease in order to demonstrate the Frenet-Serret frame, which is the same for either side, and the Darboux frames for the regions $(+)$, on the left, and $(-)$, on the right. The generators are also shown.}
\end{figure}

To establish our notation, we review the geometry of a single curved fold \cite{DuncanDuncan82,FuchsTabachnikov99}, and extend these results to recursively construct multiple pleated structures. Let $\mathbf{c}_i(s)$ be the embedding of a set of folds, labeled by $i$, in space. We choose $s$ to be the arc length of the first fold, and define $s^i$ to be the arc length of the $i^{th}$ fold. Consider the diagram in Fig. \ref{fig:multi2}{\bf a} showing three consecutive folds. The tangent planes on either side $(\pm)$ of the $i^{th}$ fold, where $(+)$ and $(-)$ respectively denote the outside and the inside sheets (Fig. \ref{fig:multi2}{\bf b}), are spanned by the unit tangent $\hat{\mathbf{t}}^i=\partial_{s^i}\mathbf{c}^i$ and its orthogonal unit vectors $\hat{\mathbf{u}}_\pm^i$. For completeness, we also define $\hat{\mathbf{N}}^i_\pm=\hat{\mathbf{t}}^i\times\hat{\mathbf{u}}_\pm^i$ to be normals to each tangent plane. Since we assume the surfaces on either side of the $i^{th}$ fold are developable, they are spanned by two unit-length generators $\hat{\mathbf{g}}_\pm^i$ respectively. The relationship between two consecutive folds is, therefore,
\begin{equation}\label{eq:embedding}
\mathbf{c}^{i}(s)=\mathbf{c}^{i-1}(s)+v _{+}^{i-1}(s)\hat{\mathbf{g}}^{i-1}_{+}(s),
\end{equation}
where
\begin{equation}
\label{eq:generator}
\hat{\mathbf{g}}^i_\pm(s)=\hat{\mathbf{t}}^i(s)\cos\left[\gamma_\pm^{i}(s)\right]\mp\hat{\mathbf{u}}^i_\pm(s)\sin\left[\gamma_\pm^{i}(s)\right],
\end{equation} 
where the variables $\gamma_\pm^i$ are the angles between the generators and unit tangent. The distance $v_{+}^{i-1}(s)$ is the distance along the generator from the $(i-1)^{th}$ fold to the $i^{th}$ fold so it depends implicitly on $\gamma_+^i$.

It can be seen from the diagram in Fig. \ref{fig:multi2}{\bf b} that the decomposition of the vectors $\hat{\mathbf{u}}_\pm^i(s)$ are as follows:
\begin{equation}
\label{eq:tanvec}
\hat{\mathbf{u}}^i_\pm(s)=\hat{\mathbf{n}}^i(s)\sin\left[\frac{\theta^{i}(s)}{2}\right]\pm\hat{\mathbf{b}}^i(s)\cos\left[\frac{\theta^{i}(s)}{2}\right],
\end{equation} 
where $\hat{\mathbf{n}}^i=\partial_{s^i}\hat{\mathbf{t}}^i/|\partial_{s^i}\hat{\mathbf{t}}^i|$ and $\hat{\mathbf{b}}^i=\hat{\mathbf{t}}^i\times\hat{\mathbf{n}}^i$ are the Frenet-Serret normal and binormal, respectively, and $\theta^i(s)$ is the dihedral angle of the fold. The triad of vectors $\{\hat{\mathbf{t}}^i,\hat{\mathbf{n}}^i,\hat{\mathbf{b}}^i\}$, Fig. \ref{fig:multi2}{\bf b}, forms natural moving frames on each $i^{th}$ fold, where curvature and torsion are respectively defined as follows, $\kappa^i(s)=\hat{\mathbf{n}}^i.\partial_{s^{i}}\hat{\mathbf{t}}^i$ and $\tau^i(s)=-\hat{\mathbf{n}}^i.\partial_{s^{i}}\hat{\mathbf{b}}^i$. The angles $\gamma_\pm^i$ can be expressed in terms of geodesic torsion, $\tau_{g_{\pm}}^i(s)=\tau^i \mp \partial_{s^i} \theta^i/2$, and normal curvature, $\kappa_{N_{\pm}}^i(s)=\mp \kappa^i \cos \left[\theta^i/2\right]$, as
\begin{equation}
\label{eq:newvar}
\cot\left[\gamma_\pm^{i}(s)\right]=\pm\frac{\tau_{g_\pm}^{i}(s)}{\kappa_{N_\pm}^{i}(s)}.
\end{equation}
From these definitions, it follows that $\tau_{g_+}^i=\tau_{g_-}^i - \partial_{s^i} \theta^i$ and $\kappa_{N_+}^i= - \kappa_{N_-}^i$. It is convenient to define the variable $\eta_\pm^{i}(s)\equiv\cot\left[\gamma_\pm^{i}(s)\right]$, such that
\begin{equation}
\label{eq:eta}
\eta_{+}^{i}(s)=\eta_{-}^{i}(s)+\frac{\partial_{s^{i}}\theta^{i}(s)}{\kappa^{i}_g\cot\left[\theta^{i}(s)/2\right]}
\end{equation} 
relates any two sides, $(+)$ and $(-)$, across a fold. Yet another curvature that needs a definition is the geodesic curvature, $\kappa_g^i=\kappa^i \sin \left[\theta^i/2\right]$. This quantity is the projection of the space curvature, $\kappa^i$, onto the tangent plane and can be interpreted as the curvature of the crease pattern before folding \cite{FuchsTabachnikov99}. It is important to notice that $\kappa_g^i$ is constant in $s^i$, for every $i$, if the crease pattern is made of concentric circles, similar to the patterns that generated the structures in Fig. \ref{fig:multi}. 

Fig. \ref{fig:multi2}{\bf a} shows the schematics of the fold geometry, which is constructed so that the generators of consecutive curves are related to each other by
\begin{equation}
\label{eq:genconst}
\hat{\mathbf{g}}_{+}^{i-1}(s)+\hat{\mathbf{g}}_{-}^{i}(s)=0.
\end{equation} 
When the crease pattern is given by a sequence of concentric circles evenly spaced with separation $w$, simple planar geometry allows us to write an expression for $v_\pm^{i}$ for the distance to the $(i\pm1)^{th}$ fold from the $i^{th}$,
\begin{equation}
\label{eq:vmax}
\kappa_g^iv^i_\pm=\mp \sin \gamma^i_\pm\pm \sqrt{\kappa_g^iw\left(\kappa^i_gw \pm 2\right) +\sin^2 \gamma^i_\pm},
\end{equation} 
where we have suppressed the $s$ dependence for brevity. Finally, inextensibility requires
\begin{equation}
\label{eq:kappag}
\kappa^{i}_g=\frac{\kappa^{i-1}_g}{1+w\kappa^{i-1}_g},
\end{equation}  
which resembles our choice to prescribe folds such that $\kappa^i_g<\kappa^{i-1}_g< \cdots <\kappa^1_g$.
These equations provide a means of starting from a single fold, say at $\mathbf{c}^1$, and generating the shape of all subsequent folds. First, we define the parameter $l_i(s)$ such that $\partial_{s^i} = \partial_s/l^i$. Then requiring that $\partial_{s^i} \mathbf{c}^i$ be normalized yields
\begin{eqnarray}\label{eq:l}
l^{i}&=&\left(1+w \kappa^{i-1}_g\right)l^{i-1}\nonumber\\ &&\times\left(1+\frac{\left(\partial_{s^{i-1}}\eta _{+}^{i-1}\right)v _{+}^{i-1}/\sqrt{1+\left(\eta _{+}^{i-1}\right)^2}}{1+\kappa^{i-1}_gv _{+}^{i-1}\sqrt{1+\left(\eta _{+}^{i-1}\right)^2}}\right).
\end{eqnarray}
Using the recursion relations for the embedding of the curves, equation (\ref{eq:embedding}), and the explicitly form of the rate of rotation of the frame $\{\hat{\mathbf{t}}^i,\hat{\mathbf{u}}^i_\pm,\hat{\mathbf{N}}^i_\pm\}$ (Fig. \ref{fig:multi2}{\bf b}), 
\begin{equation}
\label{eq:darboux}
\partial_{s^{i}}\left(\begin{array}{c}\hat{\mathbf{t}}^i \\ \hat{\mathbf{u}}^i_\pm \\ \hat{\mathbf{N}}^i_\pm \end{array}\right)=\left(\begin{array}{ccc}0&\kappa^i_g&\kappa_{N_\pm}^i \\ -\kappa_g^i&0&\tau^i_{g_\pm} \\ -\kappa^i_{N_\pm} &-\tau^i_{g_\pm} &0\end{array}\right)\left(\begin{array}{c}\hat{\mathbf{t}}^i \\ \hat{\mathbf{u}}^i_\pm \\ \hat{\mathbf{N}}^i_\pm \end{array}\right),
\end{equation} 
we can derive the following relationships
\begin{equation}\label{eq:kappan}
\kappa _{N_{-}}^{i}=\frac{l^{i-1}}{l^{i}}\,\frac{1+\kappa^{i-1}_g v _{+}^{i-1}\sqrt{1+\left(\eta _{+}^{i-1}\right)^2}}{1+w\kappa^{i-1} _g}\,\kappa _{N_{+}}^{i-1}
\end{equation} 
and 
\begin{equation}\label{eq:taug}
\tau _{g_{-}}^{i}=\frac{l^{i-1}}{l^{i}}\,\frac{1}{1+w\kappa^{i-1} _g}\,\tau _{g_{+}}^{i-1}.
\end{equation}
Then, the equation (\ref{eq:newvar})  yields 
\begin{equation} 
\eta_{-}^{i}=\frac{-\eta_{+}^{i-1}}{1+\kappa^{i-1}_g v_{+}^{i-1}\sqrt{1+\left(\eta _{+}^{i-1}\right)^2}}.
\end{equation} 
Finally, we have that $\theta^{i}=2\cot^{-1}\left(\kappa_{N_{-}}^{i}/\kappa^{i}_g\right)$. 

Eqs. (\ref{eq:l}), (\ref{eq:kappan}), and (\ref{eq:taug}) can be solved numerically for either closed or open fold patterns, as shown in Fig. \ref{fig:multi}{\bf a-b}. However, each subsequent fold requires higher numerical precision, so that structures with more than five folds become prohibitively difficult to evaluate computationally.

It has been an open question whether closed folds of the type shown in Fig. \ref{fig:multi}{\bf a} can exist without stretching beyond a single fold \cite{Demaine11}. The numerical evidence depicted in Fig. \ref{fig:multi}{\bf a} suggests the existence of structures, at least for a finite number of consecutive folds. To go beyond this, one must understand what kinds of obstructions can occur. Since the surfaces between folds are developable \cite{Spivak1999}, these obstructions arise as singularities because the generators cross. At these singularities, an elastic sheet will stretch at great energetic cost to avoid a diverging bending energy. In our notation, they can be expressed as bounds on how quickly the $\gamma_\pm^{i}$ can change \cite{Dias2012}:
\begin{subequations}
\label{eq:bound}
\begin{align}
\label{eq:bound1}
\gamma^{i\,\prime}_{-} &<  \frac{\sin\gamma^i_{-}}{v^i_{-}}-\kappa^{i}_g\\
\label{eq:bound2}
\gamma^{i\,\prime}_{+} &>  -\frac{\sin\gamma^i_{+}}{v^i_{+}}+\kappa^{i}_g.
\end{align}
\end{subequations}
Additional constraints arise because generators on the inside sheet of each fold may emerge from the crease at a value of $\gamma^i_-$ sufficiently small that the generator fails to meet the previous fold. This geometric constraint, which occurs when the discriminant in equation (\ref{eq:vmax}) is negative, can be translated into a bound for the torsion, 
\begin{equation}
\label{eq:bound3}
\frac{1}{\kappa^i_g}\left|\tau^i+\frac{\theta^{i\,\prime}}{2}\right| <  \frac{1-w\kappa^i_g }{\sqrt{2w\kappa^i_g-w^2\kappa^{i\,2}_g } }\cot\left(\frac{\theta^i}{2}\right).
\end{equation} 

\section{Mechanics of Helical folds}
We can find solutions to the recursion relations for open folds of constant curvature and torsion. Because constant curvature implies a constant dihedral angle, the equation (\ref{eq:eta}) becomes $\eta_+^i = \eta_-^i$. The recursion relations (\ref{eq:l}), (\ref{eq:kappan}), and (\ref{eq:taug}) then simplify to the following relations 
\begin{subequations}
\begin{align}
l^i& = \left(1+w \kappa_g^i \right) l^{i-1} \\
\kappa _{N_{-}}^{i}&=\frac{1+\kappa^{i-1}_g v _{+}^{i-1}\sqrt{1+\left(\eta _{+}^{i-1}\right)^2}}{\left(1+w\kappa^{i-1} _g\right){}^2}\kappa _{N_{+}}^{i-1}\\
\tau _{g_{-}}^{i}&=\frac{1}{\left(1+w \kappa^{i-1}_g\right){}^2}\tau _{g_{+}}^{i-1}.
\end{align}
\end{subequations}
In Fig. \ref{fig:multi}{\bf b} we show a solution for three consecutive folds generated by an initial open helix. 
A singularity occurs on the $n^{th}$ fold when
\begin{equation}
\sqrt{(r+n w)} |\tau_n| = \frac{1-w/(r+n w)}{\sqrt{2 - w/(r+n w)}} \cot \left(\frac{\theta}{2} \right).
\end{equation}
As long as these bounds are always satisfied, there is a solution to the recursion relations corresponding to a curved fold. In fact, it appears that the constraints can always be satisfied for sufficiently small $w$. Since $v_\pm^i \approx w/\sin \gamma_\pm^i$ for $\kappa_g^i w \ll 1$, $w$ can be decreased so that the right hand sides of Eqs. (\ref{eq:bound}) and (\ref{eq:bound3}) become arbitrarily large on folds adjacent to the $i^{th}$ fold. Whether the structure must terminate after a finite number of folds for any particular spacing $w$, of course, is more difficult to determine.

\begin{figure*}
\centering
\includegraphics[width=5in]{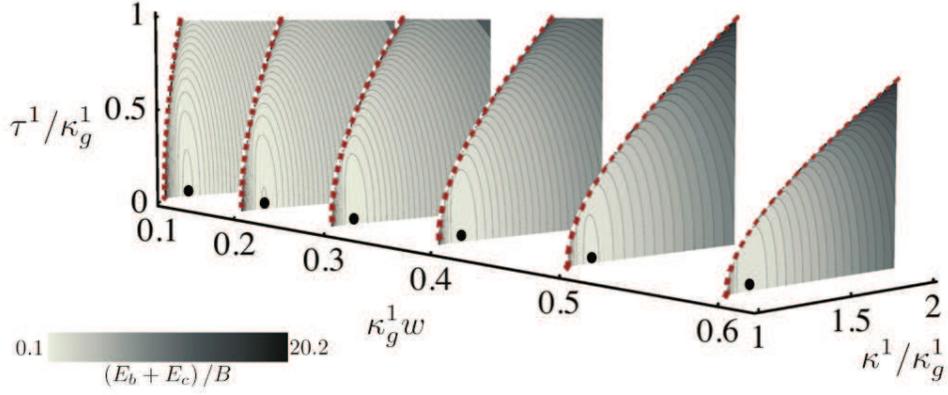}
\caption{\label{fig:MultiEnergy}  Energy landscapes normalized by $B$. We have set $\varsigma=1$ and $\theta^i_0=2\pi/3$ for all $i\in\{1,2,...,n\}$. The black dots on each energy landscape represents the energy minima at $\tau^1/\kappa_g^1=0$ and $\kappa^1/\kappa_g^1=1.15,\,\,1.14,\,\,1.13,\,\,1.12,\,\,1.11,\,\,\mbox{and}\,\,1.1$ respectively for $\kappa_g^1w=0.1,\,\,0.2,\,\,0.3,\,\,0.4,\,\,0.5,\,\,\mbox{and}\,\,0.6$. The red dashed lines mark the limit of the singular region determined by Eqn. (\ref{eq:bound3}).}
\end{figure*}

We formulate the mechanics of these structures by adding the individual contributions of bending energies coming from each pleat as well as fold energies coming from each crease $i$. The developable sheets on either side of each fold have bending energy \cite{Dias2012}
\begin{eqnarray}
\label{eq:bending}
E^{i}_{b\,\pm}&=&\frac{B}{8} \int_0^{L} \!\!ds\,l^{i}\frac{\left(\kappa^{i}_{N\pm}\right)^{2}\csc^2\gamma^{i}_{\pm}}{\kappa^{i}_g\mp\gamma^{i\,\prime}_\pm}\\ \nonumber
& & \times \ln\left(\frac{\sin\gamma^{i}_{\pm}}{\sin\gamma^{i}_{\pm}-v^{i}_{\pm}\left(\kappa^{i}_g\mp\gamma_\pm^{i\,\prime}\right)}\right),
\end{eqnarray} 
where $B$ is the bending stiffness. For a multiply folded structure, we perform a constrained sum over adjacent pairs of folds in order to avoid double counting facets when summing over $i$. Therefore, the energy for a structure with $n$ folds is given by
\begin{eqnarray}
\label{eq:benging2}
E_{b}=\left\{
\begin{array}{ll}
\sum\limits_{i=1}^{(n+1)/2} {\sum\limits_{\pm}}\,E^{2 i-1}_{b\,\pm}&\mbox{for $n$ odd}\\\\
E^{n}_{b\,+}+\sum\limits_{i=1}^{(n-1)/2}{\sum\limits_{\pm}}\,E^{2 i -1}_{b\,\pm}&\mbox{for $n$ even}.
\end{array}\right.
\end{eqnarray}
As in \cite{Dias2012}, we use
\begin{equation}
\label{eq:phenom}
E_{c}=\frac{B\varsigma}{2}\sum_{i}^n\int_0^{L} ds\,l^{i}\left[\cos\left(\frac{\theta^{i}}{2}\right)-\cos\left(\frac{\theta^{i}_0}{2}\right)\right]^2,
\end{equation} 
for the fold energies, where $\varsigma$ is a measure of the stiffness of the fold and $\theta^{i}_0$ is the preferred angle for each crease.

Typical energy landscapes are shown in Fig. \ref{fig:MultiEnergy}. These were calculated for three fold structures, like the one shown in Fig. \ref{fig:multi}{\bf b} using the formulas (\ref{eq:benging2}) and (\ref{eq:phenom}). Singularities appear along the dashed curve in Fig. \ref{fig:MultiEnergy}, as given by the relation (\ref{eq:bound3}). Beyond this limit, stretching becomes necessary in order to avoid a diverging bending energy. Finally,  the minima of energy all have $\tau^i=0$, as can be verified by folding paper.

\section{Continuum Limit}

Artistically, origami is used as a medium for sculpture. In that sense, one feels a distinct impression that a folded sheet approximates a smooth shape. An origami structure made from many nested squares, for example, has the appearance of a saddle and, in fact, asymptotically approaches a hyperbolic paraboloid (when extra creases are added) as the number of folds becomes large \cite{Demaine11}. A structure with concentric, circular folds also has the impression of a discrete approximation of a smooth saddle (Fig. \ref{fig:multi}{\bf a}) and, in this section, we seek equations to describe the resulting smooth shape.

\begin{figure}
\centering
\includegraphics[width=2in]{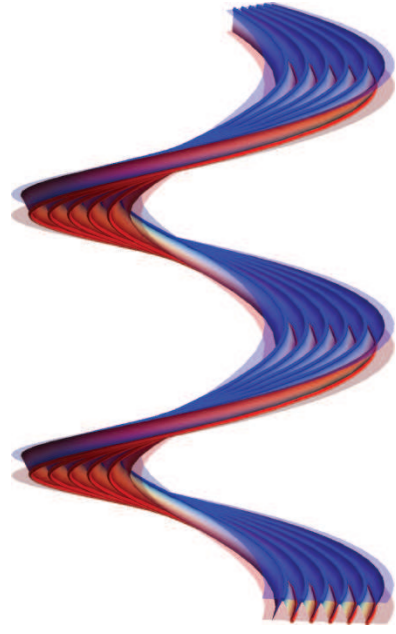}
\caption{\label{fig:approxsurf} {Helicoid surface approximating helical pleated structure. The translucent surfaces are the solutions of the continuum limit $\mathbf{c}(s,w)$, for $\kappa_0=2$ and $\tau_0=1$. The top (blue) surface touches the mountains of the pleated surface, while the bottom surface (red) touches the valleys.}}
\end{figure}

To be more precise, consider a smooth surface passing through all of the mountain folds of an origami sculpture, as shown in figure Fig. \ref{fig:approxsurf}. We seek equations governing the shape of this surface as the width between folds decreases but the number of folds diverges. To proceed, we consider two mountain folds and the valley fold lying between them. Since $w$ is small, we only express results up to $\mathcal{O}\left(w\right)$. Then
\begin{equation}
\label{eq:difference2}
\frac{\mathbf{c}^{i+1} - \mathbf{c}^{i-1}}{2w} = \frac{\partial_{s^{i}}\theta^{i}}{2\kappa^{i}_g\cot\left(\theta^{i}/2\right)}\hat{\mathbf{t}}^i-\sin\left(\frac{\theta^{i}}{2}\right)\hat{\mathbf{n}}^i+\mathcal{O}\left(w\right).
\end{equation}
Taking the limit where $w\rightarrow0$, we let $i$ become a continuous independent variable, denoted $\chi$. This yields a continuum approximation for the origami structure that can be interpreted as the initial mountain (or valley) crease evolving with a ``velocity''  defined by 
\begin{equation}
\label{eq:velocity}
\mathbf{V}(s,\chi)\equiv\partial_\chi \mathbf{c}(s,\chi) = \alpha(s,\chi) \hat{\mathbf{t}}(s,\chi) + \beta(s,\chi) \hat{\mathbf{n}}(s,\chi),
\end{equation} 
where the collection of all ``snapshots'' along the variable $\chi$ traces out a continuous surface. Here we define the components of $\mathbf{V}(s,\chi)$ as 
\begin{eqnarray}\label{eq:velocities}
\alpha(s,\chi)&\equiv&\frac{\partial_s\theta(s,\chi)}{2\,l(s,\chi)\kappa_g(\chi)\cot\left[\theta(s,\chi)/2\right]}\\ 
\beta(s,\chi)&\equiv&-\sin\left[\frac{\theta(s,\chi)}{2}\right]\end{eqnarray}
These components can be entirely expressed in terms of the curvatures through the constraint $\theta=\sin^{-1}\left(\kappa_g/\kappa\right)$. Since the discrete variation (\ref{eq:difference2}) and the velocity (\ref{eq:velocity}) are spanned by $\{\hat{\mathbf{t}},\hat{\mathbf{n}}\}$, we note that they both lie on the osculating plane of each curve. Therefore, the osculating plane becomes the tangent plane to the approximate surface when the limit $w\rightarrow0$ is taken. 

The above discrete equations, (\ref{eq:kappag}), (\ref{eq:l}), (\ref{eq:kappan}), and (\ref{eq:taug}), turn into continuous evolution equations for $\kappa_g(\chi)$,  $l(s,\chi)$, $\kappa(s,\chi)$ and $\tau(s,\chi)$ of the mountain folds in the continuum limit,
\begin{subequations}
\label{eq:contevol}
\begin{align}
\partial_\chi\kappa_g=-\kappa_g^2\\ 
\partial_\chi l= \partial_{s}\alpha-l\kappa\beta \\ 
\partial_\chi\kappa=\frac{1}{l}\partial_{s}\left(\frac{1}{l}\partial_{s}\beta\right)+\left(\kappa^2-\tau^2\right)\beta+\frac{\alpha}{l}\partial_{s}\kappa\\ 
\partial_\chi\tau= \frac{1}{l}\partial_{s}\left[\frac{\beta}{\kappa}\frac{1}{l}\partial_{s}\tau+2\frac{\tau}{\kappa}\frac{1}{l}\partial_{s}\beta\right] +2\kappa\tau\beta-\frac{\alpha}{l}\partial_{s}\tau.
\end{align}
\end{subequations}
The above set of equations (\ref{eq:contevol}), with appropriate initial data, forms a complete set of PDE's. Before looking for specific solutions, we establish some general formulas for the geometrical quantities on the smooth, approximating surface. We compute the surface's first fundamental form to be
\begin{equation}
\textrm{I} = l^2\,\, ds^2 + 2 l \alpha\,\, ds d\chi + (\alpha^2 + \beta^2)\,\, d\chi^2,
\end{equation}
and second fundamental form
\begin{equation}
\textrm{II} = -2 l \beta \tau\,\, ds d\chi + \left[-2 \alpha \beta \tau-\frac{\beta}{l \kappa} \left(\partial_s \beta \tau + \tau \partial_s \beta\right)\right] \,\,d\chi^2.
\end{equation}
Thus, the mean curvature is
\begin{equation}
H = \frac{2 \tau \partial_s \kappa - \kappa \partial_s \tau}{2 l \kappa^2},
\end{equation}
and Gaussian curvature 
\begin{equation}
K = - \tau^2.
\end{equation} 
This final result implies that $K \le 0$, with $K=0$ only for a flat surface.

When $\kappa$ and $\tau$ are independent of $s$, $\partial_\chi l = l/\chi$ so that $l = \chi$, where we have chosen $s$ to be arc length when $\chi=1$. Thus,
\begin{equation}
\partial_\chi \tau = -2 \frac{\tau}{\chi},
\end{equation}
so that $\tau = \tau_0/\chi^2$. Finally, this implies that
\begin{equation}
\partial_\chi \kappa = -\frac{\kappa^2 - \tau^2}{\kappa\chi}.
\end{equation}
This equation has solution $\kappa = \sqrt{\left(\kappa_0^2+\tau_0^2\right)\chi^{2}-\tau_0^2}/\chi^2$.
Notice that $H=0$ and $K = - \tau^2 = -\tau_0^2/\chi^4$ varies only along $\chi$. This information is already sufficient to determine that the shape must be a member of the helicoid-catenoid family. In fact, it can be verified that the surface embedding is given by 
\begin{equation}
\label{eq:emb}
\mathbf{c}(s,\chi)=\frac{\chi^2}{{\kappa _0^2+\tau _0^2}}\left(
\begin{array}{c}
\kappa \cos \left(s \sqrt{\kappa _0^2+\tau_0^2}\right)\\
\kappa\sin \left(s \sqrt{\kappa _0^2+\tau _0^2}\right) \\
\tau s\sqrt{\kappa _0^2+\tau _0^2}
\end{array}
\right), 
\end{equation}
which is indeed a helicoid. Fig. \ref{fig:approxsurf} shows how well the solutions (\ref{eq:emb}) approximate the corrugated surface.

The helicoid solution suggests that, in order to obtain the asymptotic behavior of more complex structures, one should use the following rescalings for $l(s,\chi)$, $\kappa(s,\chi)$ and $\tau(s,\chi)$: $\bar{l} = l/\chi$, $\bar{\kappa} = \chi \kappa$ and $\bar{\tau} = \tau \chi^2$. Thus, we have $\alpha = \frac{1}{\bar{l}} \partial_s (\theta/2) \tan (\theta/2)$ and the evolution equations
\begin{subequations}
\begin{align}
\partial_\chi \bar{l} = \frac{1}{\chi} \partial_s \alpha\\
\partial_\chi \bar{\tau} = -\frac{\alpha}{\chi} \frac{\partial_s \bar{\tau}}{\bar{l}}+ \frac{1}{\bar{l}} \partial_s \left[ \frac{1}{\chi \bar{\kappa}^2} \frac{\partial_s \bar{\tau}}{\bar{l}} - \frac{\bar{\tau} \partial_s \bar{\kappa}}{\chi^2 \bar{\kappa}^2 \bar{l}} \right]\\
\partial_\chi \bar{\kappa} = -\frac{1}{\bar{l}} \partial_s \left[ \frac{1}{\bar{l}} \partial_s \left(\frac{1}{\bar{\kappa}} \right) \right] + \frac{\alpha}{\chi} \frac{\partial_s \bar{\kappa}}{\bar{l}} - \frac{1}{\chi} \frac{\bar{\tau}^2}{\bar{\kappa}}
\end{align}
\end{subequations}
As $\chi$ becomes large, these equations reduce to the asymptotic equations 
\begin{subequations}
\begin{align}
l = \chi\\ 
\tau = \tau_0(s)/\chi^2\\
\partial_\chi \bar{\kappa} \approx -\partial_s^2 (1/\bar{\kappa}).
\end{align}
\end{subequations}  
Writing $\Gamma \equiv 1/\bar{\kappa}$, we see that this last equation is a nonlinear diffusion equation of the form $\partial_\chi \Gamma = \Gamma^2 \partial_s^2 \Gamma$. In the large $\chi$ limit, $\Gamma$ will become $\chi$-independent, implying that there is a $\kappa_0(s)$ such that it satisfies $-\partial_s \kappa_0/\kappa_0 = C$ asymptotically, where $C$ is a constant number. The only periodic solution is $\kappa_0$ equal to a constant. This means that initial curvatures flow towards a fixed point, the constant value $\kappa_0$. Interestingly, this scaling form for the curvature implies that $\theta$ becomes nearly constant in $s$ and $\chi$ as $\chi$ becomes large. This satisfying result indicates that concentric, circular folds are not overly frustrated from the point of view of their fold angles. In particular, we expect $\theta$ to be very close to its prescribed value, even when the fold angle cannot be exactly constant because of geometrical constraints. The mean curvature of the asymptotic surface is
\begin{equation}
H(s) = - \frac{1}{\chi^2} \frac{\partial_s \tau_0(s)}{2 \kappa_0}.
\end{equation}

\section{Conclusion}

In this article we explored the question of designing pleated folds. We were particularly interested in concentric circles crease patterns which are folded by alternating mountains and valleys. We provided the construction of two examples, closed folds that yield a saddle-like shape and open ones resulting in helical shapes. Closed folds are particularly interesting objects because once they are creased along closed paths they are in a frustrated state \cite{Dias2012}. Therefore, in order to balance their internal forces, they undergo buckling which leads to self-folding due to its highly constraint geometry. This suggests that mechanics should be fundamental to determine the equilibrium configurations of these structures. Although Demaine \emph{et al.}  \cite{Demaine11} showed that a fold pattern of concentric squares cannot be rigidly folded without either adding additional folds or internally stretching the paper, the question of concentric circles is still open. Our numerical and analytic results strongly suggest that the conjecture of \cite{Demaine11} that concentric circles can be folded without stretching is, indeed, true. A full proof, however, would need to show that the constraints of Eqs. (\ref{eq:bound}) and (\ref{eq:bound3}) do not occur for sufficiently small but finite $w$, not just when $w$ approaches zero.

\acknowledgements
We are grateful to Bryan Gin-ge Chen for the useful discussions. We thank the referees for careful reading of the manuscript. We also thank the following funds, NSF DMR 0846582, NSF-supported MRSEC on Polymers at UMass (DMR-0820506), and NSF EFRI ODISSEI-1240441.

\end{document}